\documentclass[aps,prd,showpacs,onecolumn,preprintnumbers,notitlepage,
nofootinbib,amsfont,amssymb,10pt,singlespacing] {revtex4-1}
\usepackage{amsmath,bm}
\usepackage{times}
\usepackage{braket}
\usepackage{color,graphicx}
\usepackage{slashed}
\usepackage{CJK,upgreek,fancyhdr}
\usepackage{hyperref}

\newcommand{\beq}{\begin{eqnarray}}
\newcommand{\eeq}{\end{eqnarray}}

\def\lsim{ {\ \lower-1.2pt\vbox{\hbox{\rlap{$<$}\lower6pt\vbox{\hbox{$\sim$}
}}}\ } }
\def\gsim{ {\ \lower-1.2pt\vbox{\hbox{\rlap{$>$}\lower6pt\vbox{\hbox{$\sim$}
}}}\ } }

\def \npb{  Nucl. Phys. B }

\def \plb{  Phys. Lett. B }

\def \prd{  Phys. Rev. D }
\def \prl{  Phys. Rev. Lett.  }

\def \jhep{ J. High Energy Phys.  }

\definecolor{Red}{rgb}{1.,0.,0.}

\definecolor{Blue}{rgb}{0.,0.,1.}

\newcommand{\orcid}[1]{\thanks{\href{http://orcid.org/#1}{ORCID: #1}}}

\definecolor{nicered}{rgb}{0.7,0.1,0.2}
\definecolor{nicegreen}{rgb}{0.1,0.4,0.2}

\hypersetup{colorlinks,citecolor=nicegreen,linkcolor=nicered}

\begin{document}
\begin{CJK*}{GB}{gbsn}
\title{\boldmath Studies on the $B \to \kappa \bar \kappa$ decays in the perturbative QCD
approach\footnote{\it Research exercises for excellent undergraduate students.}}
\author{Liangliang~Su(ËÕÁÁÁÁ)}
\author{Zewen~Jiang(½¯ÔóÎÄ)}
\author{Xin~Liu(ÁõÐÂ)}
\email
[ Corresponding author: ]
{liuxin@jsnu.edu.cn}
\orcid{ 0000-0001-9419-7462}
\affiliation{School of Physics and Electronic Engineering,
Jiangsu Normal University, Xuzhou 221116, China}


\date{\today}

\begin{abstract}

The $B \to \kappa \bar \kappa$ decays are investigated for the first time
in the perturbative QCD formalism based on the $k_T$ factorization theorem,
where the light scalar $\kappa$ is assumed as a two-quark state.
Our numerical results and phenomenological analyses on the
{\it CP}-averaged branching ratios and {\it CP}-violating asymmetries
show that:
(a) the $B_s^0 \to \kappa^+ \kappa^-$ and $B_s^0 \to \kappa^0 \bar \kappa^0$
decays have large decay rates around
${\cal O}(10^{-5})$, which could be examined by
the upgraded Large Hadron Collider beauty and/or Belle-II experiments in the near future;
(b) a large decay rate about $3 \times 10^{-6}$ appears in the pure annihilation
$B_d^0 \to \kappa^+ \kappa^-$ channel, which could provide more evidences to help
distinguish different QCD-inspired factorization approaches, even understand the annihilation decay mechanism;
(c) the pure penguin modes $B_d^0 \to \kappa^0 \bar \kappa^0$ and $B_s^0 \to \kappa^0
\bar \kappa^0$ would provide a promising ground
to search for the possible new physics
because of their zero direct and mixing-induced {\it CP} violations
in the standard model.
The examinations with good precision from the future experiments
will help to further study the perturbative and/or nonperturbative
QCD dynamics involved in these considered decay modes.

\end{abstract}


\preprint{\footnotesize JSNU-HEP-2019-1}
\maketitle

In the conventional quark model, a meson is composed of one quark and one
antiquark, i.e., $q\bar q$, with different coupling of the orbital and spin angular momenta~\cite{GellMann:1964nj,Zweig:1981pd,Zweig:1964jf}.
To date, the structure of the $S$-wave ground state mesons has almost
been determined unambiguously, though the $\eta$ and $\eta^\prime$
ones may contain the component of gluonium( or pseudoscalar glueball) with different
extent~\cite{Kou:1999tt,Cheng:2008ss,Liu:2012ib}.
However, the components of the $P$-wave mesons are not easily determined. In particular,
the description of the inner structure for the light scalar states such as $a_0(980)$,
$\kappa$ or $K_0^*(800)$, $\sigma$ or $f_0(500)$, and $f_0(980)$ is controversial, e.g., $q \bar q$, $\bar q\bar q q q$,
meson-meson bound states, etc., and still not well established currently(for a review,
see e.g., Refs.~\cite{Godfrey:1998pd,Close:2002zu,Tanabashi:2018oca}). When the light
scalar $f_0(980)$ was first observed in the $B \to f_0(980) K$ channel, performed by the
Belle~\cite{Abe:2002av} and BABAR~\cite{Aubert:2003mi} collaborations in 2002 and 2004,
respectively, the investigations on the
light scalars in the decay productions of the heavy $B$ mesons were naturally considered
as a unique insight to explore their underlying structure. With many channels including light scalars of the heavy $B$ meson decays being opened experimentally~\cite{Amhis:2016xyh,Tanabashi:2018oca}, $B \to SP, SV$
(Here, $P$ and $V$ denote the pseudoscalar
and vector meson, respectively) decays have been
studied extensively at the theoretical aspects with different
approaches/methods, for instance,
see~\cite{Cheng:2009xz,Liu:2009xm,Liu:2010zg,Liu:2013cvx,Shen:2006ms,Wang:2006ria,Wang:2009azc,Colangelo:2010bg,Kim:2009dg}. With the great development of the Large Hadron Collider beauty(LHCb) and Belle-II experiments~\cite{Kou:2018nap}, more and more modes
involving one and/or two scalar states in the $B$ meson decays are expected to be measured
with good precision in the future.

In this work, we will study the charmless hadronic $B \to \kappa \bar \kappa$ decays
(Here, $B$ denotes the nonstrange $B^+$ and $B_{d}^0$, and strange $B_s^0$ mesons.)
for the first time by employing the perturbative QCD(PQCD)
approach~\cite{Keum:2000wi,Lu:2000em,Lu:2000hj} based on the $k_T$ factorization theorem,
where the light scalar $\kappa$ will be considered as a lowest-lying $q \bar q$ state.
Theoretically, the most important part of a nonleptonic decay
amplitude is the effective calculation
of the hadronic matrix element, in which the essential inputs are the wave functions
(or light-cone distribution amplitudes) of the initial and final hadron states that
describe the nonperturbative QCD dynamics independent on
the processes. The PQCD approach, as one of the presently three
popular QCD-inspired factorizations (the other two are QCD factorization approach~\cite{Beneke99:qcdf,Du02:qcdf}
and soft-collinear effective theory~\cite{Bauer04:scet}, respectively),
has the advantages in computing the Feynman amplitudes by conquering
the endpoint singularities that exist in the collinear factorization
theorem. By keeping the transverse momentum of the valence quark,
associated with the Sudakov factors arising from the $k_T$ resummation~\cite{Botts89:ktfact,Li92:sudakov} and threshold  resummation~\cite{Li02:threshold}, the PQCD approach can be well
applied to calculate the hadronic matrix
element of the nonleptonic $B$ meson decays. Apart from the
factorizable emission diagrams, the
nonfactorizable emission ones and the annihilation ones can
also be perturbatively calculated.
Furthermore, even though the origin of the {\it CP} violation
and the annihilation decay mechanism are
currently unknown, the experimental measurements~\cite{Amhis:2016xyh,Tanabashi:2018oca} performed by the BABAR, Belle, and LHCb collaborations
have confirmed the direct {\it CP}-violating asymmetry of
the $B \to K\pi$ decays~\cite{Keum:2000wi,Keum:2000ph} and the
large decay rates of the pure annihilation $B_d^0 \to
K^+ K^-$ and $B_s^0 \to \pi^+ \pi^-$ modes~\cite{Li:2004ep,Xiao:2011tx}
predicted in the PQCD approach. Certainly,
the predictions made in the PQCD approach about the branching ratios and {\it CP} violations of the $B \to PP, PV/VP,$ and $VV$ decays generally agree with the
available data within errors.

At the quark level, the considered $B \to \kappa \bar \kappa$ decays are
induced by the $\bar b \to \bar d$
or $\bar b \to \bar s$ transitions, respectively. The weak effective Hamiltonian $H_{\rm eff}$ for the $B \to \kappa \bar  \kappa$ decays can be written as~\cite{Buchalla:1995vs},
\begin{equation}
H_{\rm eff}\, =\, {G_F\over\sqrt{2}}
\left\{V_{ub}^*V_{uQ} \left[C_1(\mu)O_1^{u}(\mu)
+C_2(\mu)O_2^{u}(\mu)\right]- V_{tb}^*V_{tQ} \sum_{i=3}^{10}C_i(\mu)O_i(\mu)\right\}\;,
\label{eq:heff}
\end{equation}
with the Fermi constant $G_F=1.16639\times 10^{-5}{\rm GeV}^{-2}$, the light $Q = d, s$ quark, and Wilson
coefficients $C_i(\mu)$ at the renormalization scale
$\mu$. The local four-quark operators $O_i(i=1,\cdots,10)$ are written as
\begin{itemize}
\item{ current-current(tree) operators}
\begin{eqnarray}
{\renewcommand\arraystretch{1.5}
\begin{array}{ll}
\displaystyle
O_1^{u}\, =\,
(\bar{Q}_\alpha u_\beta)_{V-A}(\bar{u}_\beta b_\alpha)_{V-A}\;,
& \displaystyle
O_2^{u}\, =\, (\bar{Q}_\alpha u_\alpha)_{V-A}(\bar{u}_\beta b_\beta)_{V-A}\;;
\end{array}}
\label{eq:operators-1}
\end{eqnarray}

\item{ QCD penguin operators}
\begin{eqnarray}
{\renewcommand\arraystretch{1.5}
\begin{array}{ll}
\displaystyle
O_3\, =\, (\bar{Q}_\alpha b_\alpha)_{V-A}\sum_{q'}(\bar{q}'_\beta q'_\beta)_{V-A}\;,
& \displaystyle
O_4\, =\, (\bar{Q}_\alpha b_\beta)_{V-A}\sum_{q'}(\bar{q}'_\beta q'_\alpha)_{V-A}\;,
\\
\displaystyle
O_5\, =\, (\bar{Q}_\alpha b_\alpha)_{V-A}\sum_{q'}(\bar{q}'_\beta q'_\beta)_{V+A}\;,
& \displaystyle
O_6\, =\, (\bar{Q}_\alpha b_\beta)_{V-A}\sum_{q'}(\bar{q}'_\beta q'_\alpha)_{V+A}\;;
\end{array}}
\label{eq:operators-2}
\end{eqnarray}

\item{ electroweak penguin operators}
\begin{eqnarray}
{\renewcommand\arraystretch{1.5}
\begin{array}{ll}
\displaystyle
O_7\, =\,
\frac{3}{2}(\bar{Q}_\alpha b_\alpha)_{V-A}\sum_{q'}e_{q'}(\bar{q}'_\beta q'_\beta)_{V+A}\;,
& \displaystyle
O_8\, =\,
\frac{3}{2}(\bar{Q}_\alpha b_\beta)_{V-A}\sum_{q'}e_{q'}(\bar{q}'_\beta q'_\alpha)_{V+A}\;,
\\
\displaystyle
O_9\, =\,
\frac{3}{2}(\bar{Q}_\alpha b_\alpha)_{V-A}\sum_{q'}e_{q'}(\bar{q}'_\beta q'_\beta)_{V-A}\;,
& \displaystyle
O_{10}\, =\,
\frac{3}{2}(\bar{Q}_\alpha b_\beta)_{V-A}\sum_{q'}e_{q'}(\bar{q}'_\beta q'_\alpha)_{V-A}\;.
\end{array}}
\label{eq:operators-3}
\end{eqnarray}
\end{itemize}
with the color indices $\alpha, \ \beta$ and the notations
$(\bar{q}'q')_{V\pm A} = \bar q' \gamma_\mu (1\pm \gamma_5)q'$.
The index $q'$ in the summation of the above operators runs
through $u,\;d,\;s$, $c$, and $b$. It is worth mentioning that
since we work in the leading order[${\cal O}(\alpha_s)$] of the PQCD
approach, it is consistent to use the leading order Wilson coefficients.
For the renormalization group evolution of the Wilson coefficients
from higher scale to lower scale, the formulas as given in
Refs.~\cite{Keum:2000wi,Lu:2000em} will be adopted directly.

\begin{figure}[htb]
\centering
\begin{tabular}{l}
\includegraphics[width=0.7\textwidth]{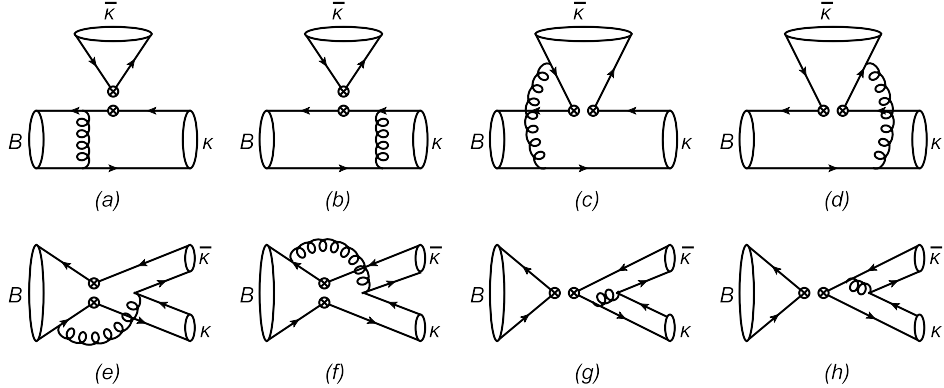}
\end{tabular}
\caption{Leading order Feynman diagrams for $B \to \kappa \bar \kappa$ decays in the PQCD formalism }
\label{fig:fig1}
\end{figure}

The Feynman diagrams of the $B \to \kappa \bar \kappa$ decays at leading order in the PQCD
formalism are illustrated in Fig.~\ref{fig:fig1}:
\begin{itemize}
\item{Emission topology}: Fig.~\ref{fig:fig1}(a) and \ref{fig:fig1}(b) describe the factorizable emission diagrams, while Fig.~\ref{fig:fig1}(c) and \ref{fig:fig1}(d) describe the nonfactorizable emission ones;
\item{ Annihilation topology}: Fig.~\ref{fig:fig1}(e) and \ref{fig:fig1}(f) describe the nonfactorizable annihilation diagrams, while Fig.~\ref{fig:fig1}(g) and \ref{fig:fig1}(h) describe the factorizable annihilation ones.
\end{itemize}
In 2013, one of us(X.~Liu) with Xiao and Zou ever studied the $B \to K_0^*(1430)
\bar K_0^*(1430)$ decays in the PQCD approach~\cite{Liu:2013lka}, where the analytic
expressions for the factorization formulas and the decay amplitudes were
presented explicitly. Therefore, we just need to
replace the $K_0^*(1430)$ state in Ref.~\cite{Liu:2013lka} with the light $\kappa$ one to
obtain easily the corresponding information of the $B \to \kappa \bar \kappa$ decays
in the PQCD approach. Hence, for simplicity, we will not collect the aforementioned formulas in this paper. The interested readers can refer to Ref.~\cite{Liu:2013lka}
for detail.

Then, we can turn to the numerical calculations of the {\it CP}-averaged
branching ratios and {\it CP}-violating asymmetries of the $B \to \kappa \bar \kappa$
decays in the PQCD approach. Before proceeding, some essential comments on the nonperturbative inputs are as follows:
\begin{enumerate}
\item[]{(a)} For the heavy $B$ mesons, the wave functions and the distribution
amplitudes, and the decay constants are same as those utilized in Ref.~\cite{Liu:2013lka}, but with the updated lifetimes $\tau_{B_d^0} = 1.52$~ps and $\tau_{B_s^0} = 1.509$~ps, which can be found clearly in the newest Review of Particle Physics~\cite{Tanabashi:2018oca}.

\item[]{(b)} For the light scalar $\kappa$, the decay constants and the Gegenbauer moments in the distribution amplitudes have been derived at the normalization scale $\mu=1$ GeV in the QCD sum rule
    method~\cite{Cheng:2005nb}: the scalar decay constant $\bar f_{\kappa} = 0.34 \pm 0.02$~GeV, the vector decay constant $f_\kappa = \bar f_{\kappa}/\mu$ with $\mu= m_{\kappa}/(m_s- m_q)$ ($m_{\kappa}$, $m_s$, and $m_q$ stand for the masses of the light scalar $\kappa$, the strange quark $s$, and the nonstrange light quark $u$ and $d$, respectively.),
    and the Gegenbauer moments $B_1= -0.92 \pm 0.11$ and $B_3= 0.15 \pm 0.09$. Here,
    the running current quark masses
     $m_s = 0.12$~GeV and $m_q = 0.005$~GeV at $\mu=1$~GeV, which are translated from those
     in a $\overline{\rm MS}$ scale $\mu \approx 2$ GeV~\cite{Tanabashi:2018oca}, are adopted in the calculations. Note that the isospin symmetry is assumed in this work. For the light scalar $\kappa$ mass $m_\kappa$, we adopt the value
     $m_\kappa = 0.8$ GeV for rough estimations, because this scalar $\kappa$ has been assumed
     as the lowest-lying $q \bar q$ state~\footnote{ Moreover, as inferred from the newest
     Review of Particle Physics~\cite{Tanabashi:2018oca}, this state is also with
     a finite but indefinite width, whose effect, in principle, has to be included
     to make relevant predictions more precise. Generally speaking, the width effect could result in the enhancement/reduction of the numerical results with different extent~\cite{Cheng:2003xc}. However, up to now, to our best knowledge,
     the essential $S$-wave $K\pi$ distribution amplitudes for resonance $\kappa$ state with the
     constrained parameters, e.g., Gegenbauer moments, are absent.
     Therefore, the width effect will be left for future
     investigations elsewhere.
     }.

\item[]{(c)} For the Cabibbo-Kobayashi-Maskawa(CKM) matrix elements, we also adopt the Wolfenstein
parametrization at leading order, but with the updated parameters $A=0.836$, $\lambda=0.22453$, $\bar \rho= 0.122^{+0.018}_{-0.017}$, and $\bar \eta= 0.355^{+0.012}_{-0.011}$~\cite{Tanabashi:2018oca}.
\end{enumerate}

Now, we present the numerical results of the $B \to \kappa \bar \kappa$ decays in the PQCD formalism. Firstly, the PQCD predictions of the {\it CP}-averaged branching ratios
can be read as follows:
\beq
Br(B^+ \to \kappa^+ \bar \kappa^0) &=& 5.46^{+0.17}_{-0.06}(\omega_B)^{+2.25+1.39}_{-1.73-0.37}(B_i)^{+1.41}_{-1.18}(\bar f_{\kappa})^{+0.13+0.22}_{-0.09-0.19}(\rm CKM) \times 10^{-7}\;;
\label{eq:br-u}
\eeq
and
\beq
Br(B_d^0 \to \kappa^+ \kappa^-) &=& 2.86^{+0.19}_{-0.22}(\omega_B)^{+1.36+0.40}_{-1.00-0.31}(B_i)^{+0.74}_{-0.62}(\bar f_{\kappa})^{+0.08+0.15}_{-0.07-0.13}(\rm CKM) \times 10^{-6}\;,
\label{eq:br-d1}\\
Br(B_d^0 \to \kappa^0 \bar \kappa^0) &=& 7.75^{+0.93}_{-1.05}(\omega_B)^{+4.85+1.53}_{-3.27-0.00}(B_i)^{+1.99}_{-1.67}(\bar f_{\kappa})^{+0.08+0.27}_{-0.07-0.27}(\rm CKM) \times 10^{-7}\;;
\label{eq:br-d2}
\eeq
and
\beq
Br(B_s^0 \to \kappa^+ \kappa^-) &=&  1.15^{+0.25}_{-0.29}(\omega_B)^{+0.83+0.39}_{-0.55-0.00}(B_i)^{+0.29}_{-0.25}(\bar f_{\kappa})^{+0.00+0.01}_{-0.00-0.01}(\rm CKM) \times 10^{-5}\;,
\label{eq:br-s1}\\
Br(B_s^0 \to \kappa^0 \bar \kappa^0) &=&  1.55^{+0.24}_{-0.27}(\omega_B)^{+1.15+0.30}_{-0.75-0.00}(B_i)^{+0.40}_{-0.33}(\bar f_{\kappa})^{+0.00+0.00}_{-0.00-0.00}(\rm CKM) \times 10^{-5}\;.
\label{eq:br-s2}
\eeq
From the Eqs.~(\ref{eq:br-u})-(\ref{eq:br-s2}), one can find the following points:
\begin{itemize}
\item[]{(a)}
The considered $B \to \kappa \bar \kappa$ decays have evidently different {\it CP}-averaged branching ratios in the PQCD approach, namely, varying from $10^{-7}$ to $10^{-5}$. Frankly speaking, these numerical results suffer from large theoretical errors mainly induced by the nonperturbative inputs, such as the shape parameter $\omega_B$ in the $B$ meson distribution amplitude, the scalar decay constant $\bar f_{\kappa}$, especially the Gegenbauer moments $B_{i}(i=1,3)$ in the leading twist light-cone distribution amplitude of $\kappa$. The uncertainties of the above mentioned parameters need to be constrained by the future precise measurements and/or Lattice QCD or QCD sum rule calculations.

\item[]{(b)}
The pure annihilation decay of $B_d^0 \to \kappa^+ \kappa^-$ has the same quark
structure as that of the measured one $B_d^0 \to K^+ K^-$, whose decay rate predicted in the PQCD approach has been confirmed by the LHCb experiments~\cite{Aaij:2012as,Aaij:2016elb}. Therefore, it is expected that
the large branching ratio of the $B_d^0 \to \kappa^+ \kappa^-$ mode given in this work could be examined in the LHCb and/or Belle-II experiments. The confirmation of
this PQCD result would provide useful hints to understand the inner structure of the
light scalar $\kappa$.

\item[]{(c)}
In light of the large $Br(B_d^0 \to \kappa^+ \kappa^-)_{\rm PQCD}$ while the small
$Br(B_d^0 \to \kappa^0 \bar \kappa^0)_{\rm PQCD}$, under the assumption of isospin symmetry, it is postulated that a significant
cancellation occurred in the $B_d^0 \to \kappa^0 \bar \kappa^0$
decay between the contributions
induced by the emission and the annihilation topologies, which, as a matter of fact,
can be found clearly from the numerical results for the factorization decay amplitudes presented in Table~\ref{tab:DecayAmps}.

\item[]{(d)}
The decay rates of the $B_s^0 \to \kappa^+ \kappa^-$ and $B_s^0 \to \kappa^0 \bar \kappa^0$ modes indicate a very small contamination induced by the tree annihilation diagrams associated with a CKM-suppressed factor $V_{us} \sim \lambda$ in the $\bar b \to \bar s$ transition. Meanwhile, relative to $V_{td} \sim A\lambda^3(1-\rho-{\it i} \eta)$ in the $\bar b \to \bar d$ transition, the CKM-enhanced factor $V_{ts} \sim A\lambda^2$ involved in these two decays finally resulted in the highly large and close branching ratios around ${\cal O}(10^{-5})$.

\item[]{(e)}
As mentioned above, because of the enhanced factor $r_{\rm CKM}=|V_{ts}/V_{td}|^2 \sim 23.6$~\cite{Tanabashi:2018oca}, the pure penguin modes $B_d^0 \to \kappa^0 \bar \kappa^0$ and $B_s^0 \to \kappa^0 \bar \kappa^0$ have significantly different decay rates, namely, the former one with
$7.75^{+5.55}_{-3.83}\times 10^{-7}$ while the latter one with $1.55^{+1.28}_{-0.86}\times 10^{-5}$, respectively, where the errors have been added in quadrature. In light of the large theoretical errors, a precise ratio of these two branching ratios would be more interested,
\beq
R_{s/d}^{00}(\kappa\bar \kappa)&=&\frac{Br(B_s^0 \to \kappa^0 \bar \kappa^0)}
{Br(B_d^0 \to \kappa^0 \bar \kappa^0)} = 20.0^{+1.7}_{-2.4}\;,
\eeq
Similarly, another two interesting ratios $R_{s/d}^{+-}(\kappa\bar \kappa)$ and $R_{s}^{00/+-}(\kappa\bar \kappa)$ could be easily obtained,
\beq
R_{s/d}^{+-}(\kappa\bar \kappa)&=&\frac{Br(B_s^0 \to \kappa^+ \bar \kappa^-)}
{Br(B_d^0 \to \kappa^+ \bar \kappa^-)} = 4.0^{+1.2}_{-1.1}\;; \\
R_{s}^{00/+-}(\kappa\bar \kappa)&=& \frac{Br(B_s^0 \to \kappa^0 \bar \kappa^0)}
{Br(B_s^0 \to \kappa^+ \bar \kappa^-)} = 1.3^{+0.1}_{-0.1}\;,
\eeq
It is clearly found that the uncertainties in the above ratios $R_{s/d}^{00}$, $R_{s/d}^{+-}$, and $R_{s}^{00/+-}$ are
significantly small because the theoretical errors resulted from the hadronic inputs have been cancelled to a great extent. These values are expected to be examined in
the future $B$-physics experiments to help further understand the involved QCD dynamics
in depth.

\item[]{(f)}
In order to understand the contributions arising from different topologies better, the numerical values for the factorization
decay amplitudes are presented explicitly in Table~\ref{tab:DecayAmps}. One can find the large nonfactorizable emission contributions and the much larger annihilation contributions in the considered $B \to \kappa \bar \kappa$ decays, especially in the two $B_s^0$ modes. The underlying reason is that the antisymmetric QCD behavior from the only odd terms in the twist-2 distribution amplitude
of the light scalar $\kappa$~\cite{Cheng:2005nb},
\beq
\phi_{\kappa}(x,\mu)&=&\frac{3}{\sqrt{6}}x(1-x)\biggl\{f_{\kappa}(\mu)+\bar
f_{\kappa}(\mu)\sum_{m=1}^\infty B_m(\mu)C^{3/2}_m(2x-1)\biggr\}\;,
\label{eq:t2-ka}
\eeq
where $f_{\kappa}(\mu)$ and $\bar f_{\kappa}(\mu)$, $B_m(\mu)$, and
$C_m^{3/2}(t)$ are the vector and scalar decay constants,
Gegenbauer moments, and Gegenbauer polynomials,
respectively, make the previously destructive interferences become the presently constructive ones between the valence-quark-radiative and valence-antiquark-radiative diagrams in the nonfactorizable emission and annihilation topologies, as
already illustrated in Fig.~\ref{fig:fig1}. It is worth mentioning
that, as can be seen in Table~\ref{tab:DecayAmps}, the annihilation diagrams play a
dominant role on both of the {\it CP}-averaged decay rates and the {\it CP}
violations of the considered $B \to \kappa \bar \kappa$ decays in this work.

\item[]{(g)}
As for the experimental measurements of the predicted large branching ratios, e.g.,
$Br(B_s^0\to \kappa^0 \bar \kappa^0)=1.55^{+1.28}_{-0.86}\times 10^{-5}$ and
$Br(B_s^0 \to \kappa^+ \kappa^-)=1.15^{+0.99}_{-0.67}\times 10^{-5}$,
we expect the LHCb and/or Belle-II experiments might measure these
channels through the Dalitz plot analysis of $B_s^0 \to (K\pi)_\kappa(K\pi)_{\bar \kappa}$.
In principle, the LHCb and Belle-II experiments have the abilities
to detect the $B$ meson decay rates with large branching ratios above
$10^{-6}$. Taking $B_s^0 \to \kappa^0 \bar \kappa^0$
mode as an example,
the decay rate ${\cal B}(\kappa^0 \to K^+ \pi^-)$ is $\frac{2}{3}$ based on the assumption of isospin symmetry in the strong interactions.
Therefore, we could obtain a branching ratio
${\rm BR}(B_s^0 \to (K^+ \pi^-)_{\kappa^0} (K^- \pi^+)_{\bar \kappa^0})
\equiv Br(B_s^0 \to \kappa^0 \bar \kappa^0)\cdot {\cal B}(\kappa^0 \to K^+ \pi^-)
\cdot {\cal B}(\bar \kappa^0 \to K^- \pi^+)
= 6.89^{+5.69}_{-3.82} \times 10^{-6}$.
We hope this large value above $10^{-6}$ could
be measured by the LHCb and/or Belle-II experiments
when the events with high statistics are collected.
Certainly, more information of the intermediate state $\kappa$ demand
the studies on the
four-body $B_s^0 \to K^+ K^- \pi^+ \pi^-$ decay armed with the $S$-wave $K\pi$
distribution amplitudes with well constrained nonperturbative
parameters for $\kappa$ from
Lattice QCD and/or experimental measurements. Unfortunately, they are absent
currently to our best knowledge theoretically and experimentally.
Therefore, this issue has to be left for future studies elsewhere. 
\end{itemize}
\begin{table}[htb]
\caption{The factorization decay amplitudes(in units of $10^{-3}$~GeV$^{3}$) of the nonleptonic $B \to \kappa \bar \kappa$ decays in the PQCD approach at leading order, where only the central values are quoted for clarifications. }
\label{tab:DecayAmps}
 \begin{center}\vspace{-0.5cm}{
\begin{tabular}[t]{c||c|c|c|c}
\hline  \hline
   Modes   &  ${\cal F}_{fe}$  & ${\cal M}_{nfe}$ &${\cal M}_{nfa}$ &${\cal F}_{fa}$ \\
\hline
 $B^+ \to \kappa^+ \bar \kappa^0$
     &$0.415 - {\it i} 0.173$
     &$-0.377 - {\it i} 0.216$
     &$-0.164 + {\it i} 0.814$
     &$-0.035 + {\it i} 0.859$
 \\
 $B_d^0 \to \kappa^+ \kappa^-$
     &$-$
     &$-$
     &$-0.597 - {\it i} 1.983$
     &$0.0008 + {\it i} 0.0005$
 \\
 $B_d^0 \to \kappa^0 \bar \kappa^0$
     &$0.415 - {\it i} 0.173$
     &$-0.377 - {\it i} 0.216$
     &$-1.049 - {\it i} 0.036$
     &$-0.055 + {\it i} 0.907$
 \\
 $B_s^0 \to \kappa^+ \kappa^-$
     &$-3.144 + {\it i} 0.377$
     &$1.130 + {\it i} 0.394$
     &$5.236 + {\it i} 1.193$
     &$1.796 - {\it i} 3.709$
 \\
 $B_s^0 \to \kappa^0 \bar \kappa^0$
     &$-3.360  $
     &$1.564 + {\it i} 1.736$
     &$5.301 + {\it i} 1.782$
     &$1.788 - {\it i} 3.695$
 \\
 \hline \hline
\end{tabular}}
\end{center}
\end{table}

Then, we will discuss the {\it CP}-violating asymmetries of the $B \to \kappa \bar \kappa$ decays in the PQCD approach.
The direct and the mixing-induced {\it CP} asymmetries ${\cal A}_{\rm dir}$ and
${\cal A}_{\rm mix}$ are collected as~\footnote{It is worth pointing out that, due to the nonzero ratio $(\Delta\Gamma/\Gamma)_{B_s^0}$
for the $B_s^0-\bar{B}_s^0$ mixing as expected in the standard model, the third {\it CP} asymmetry ${\cal A}_{\rm \Delta\Gamma_s}$ will
appear in the $B_s^0 \to \kappa \bar \kappa$ decays~\cite{Liu:2013lka}. Here, the quantity $\Delta\Gamma$ is the decay width difference of the
$B_s$ meson mass eigenstates~\cite{Beneke99:Bsmixing,Fernandez06:Bsmixing}. Moreover, the three quantities
describing the {\it CP} violations in the $B_s$ meson decays satisfy the relation: $|{\cal A}_{\rm dir}|^2+|{\cal A}_{\rm mix}|^2+|{\cal A}_{\rm \Delta\Gamma_s}|^2=1$.}
\beq
{\cal A}_{\rm dir}(B^+ \to \kappa^+ \bar \kappa^0) &=& -87.1^{+14.0}_{-7.7}(\omega_B)^{+1.9+22.3}_{-0.0-8.7}(B_i)^{+0.0}_{-0.0}(\bar f_{\kappa})^{+0.6+2.1}_{-0.4-1.8}(\rm CKM)\times 10^{-2}\;;
\eeq
and
\beq
{\cal A}_{\rm dir}(B_d^0 \to \kappa^+ \kappa^-) &=&\hspace{0.25cm} 15.4^{+0.1}_{-0.7}(\omega_B)^{+1.4+3.5}_{-1.1-5.5}(B_i)^{+0.0}_{-0.0}(\bar f_{\kappa})^{+0.1+0.8}_{-0.1-0.7}(\rm CKM)\times 10^{-2}\;, \\
{\cal A}_{\rm mix}(B_d^0 \to \kappa^+ \kappa^-) &=& -80.0^{+0.0}_{-0.3}(\omega_B)^{+1.6+2.9}_{-1.3-3.6}(B_i)^{+0.0}_{-0.0}(\bar f_{\kappa})^{+3.5+1.7}_{-3.0-1.6}(\rm CKM)\times 10^{-2}\;; \\
{\cal A}_{\rm dir}(B_d^0 \to \kappa^0 \bar \kappa^0) &\approx&\hspace{0.25cm} 0.0
\;, \\
{\cal A}_{\rm mix}(B_d^0 \to \kappa^0 \bar \kappa^0) &\approx&\hspace{0.25cm} 0.0
\;;
\eeq
and
\beq
{\cal A}_{\rm dir}(B_s^0 \to \kappa^+ \kappa^-) &=& -35.8^{+6.4}_{-9.3}(\omega_B)^{+2.7+6.0}_{-3.9-0.0}(B_i)^{+0.0}_{-0.0}(\bar f_{\kappa})^{+1.0+0.2}_{-1.1-0.2}(\rm CKM) \times 10^{-2}\;,\\
{\cal A}_{\rm mix}(B_s^0 \to \kappa^+ \kappa^-) &=&\hspace{0.25cm} 12.3^{+3.4}_{-1.5}(\omega_B)^{+2.8+12.4}_{-3.8-9.5}(B_i)^{+0.0}_{-0.0}(\bar f_{\kappa})^{+0.4+0.3}_{-0.4-0.3}(\rm CKM)\times 10^{-2}\;,\\
{\cal A}_{\Delta\Gamma_s}(B_s^0 \to \kappa^+ \kappa^-) &=&\hspace{0.25cm} 92.6^{+2.4}_{-4.8}(\omega_B)^{+0.5+2.8}_{-1.2-1.3}(B_i)^{+0.0}_{-0.0}(\bar f_{\kappa})^{+0.4+0.0}_{-0.5-0.1}(\rm CKM) \times 10^{-2}\;; \\
{\cal A}_{\rm dir}(B_s^0 \to \kappa^0 \bar \kappa^0) &=&\hspace{0.25cm} 0.0
\;,\\
{\cal A}_{\rm mix}(B_s^0 \to \kappa^0 \bar \kappa^0) &=&\hspace{0.25cm} 0.0
\;,\\
{\cal A}_{\Delta\Gamma_s}(B_s^0 \to \kappa^0 \bar \kappa^0) &=&\hspace{0.25cm} 1.0
\;.
\eeq
in which the definitions of the direct {\it CP} violation ${\cal A}_{\rm dir}$, the mixing-induced one ${\cal A}_{\rm mix}$, even the third one ${\cal A}_{\rm \Delta\Gamma_s}$ arising from the nonnegligible $(\Delta\Gamma/\Gamma)_{B_s^0}$ term are same as
those in Ref.~\cite{Liu:2013lka}. From these numerical results of the {\it CP} violations of the $B \to \kappa \bar \kappa$ decays
in the PQCD approach, some comments are in order:
\begin{itemize}
\item Generally speaking, these PQCD predictions are not sensitive to the variation of the scalar decay constant $\bar f_{\kappa}$ as shown in the above Equations. This can be deduced from the tiny vector decay constant $f_{\kappa}$ in the leading twist light-cone distribution amplitude of the scalar $\kappa$ meson(See Eq.~(\ref{eq:t2-ka}) for detail). Furthermore, both of the twist-3 light-cone distribution amplitudes of $\kappa$ are proportional to the scalar decay constant $\bar f_{\kappa}$ because of adopting the asymptotic forms for simplicity~\cite{Liu:2013lka}. Based on the definitions, the {\it CP} asymmetry is the ratio of the differences of the related branching ratios between $B \to \kappa \bar \kappa$ and $\bar B \to \bar \kappa \kappa$ modes to their corresponding summations, then the scalar decay constant $\bar f_{\kappa}$ will be cancelled naturally.

\item A large direct {\it CP} violation for the $B^+ \to \kappa^+ \bar \kappa^0$ mode can be observed, $-87.1^{+26.5}_{-11.8}\%$, which indicates that the involved penguin contributions are sizable, within large theoretical errors. While, due to the small branching ratio predicted in the PQCD approach, it might not be easily measured in the near future.

\item Both of the $B_d^0 \to \kappa^+ \kappa^-$ and $B_s^0 \to \kappa^+ \kappa^-$ channels exhibit large {\it CP}-violating asymmetries, which are expected to be measured with much more possibilities at the LHCb
    and/or Belle-II experiments because of their large decay rates, namely, $2.86^{+1.62}_{-1.25} \times 10^{-6}$
    and $1.15^{+0.99}_{-0.67}\times 10^{-5}$, where the errors have been added in quadrature too. The confirmations from the future measurements on these two modes would provide the
    evidences not only to support the assumption of the two-quark structure of the light scalar $\kappa$ in the present work, but also to help distinguish different factorization approaches on clarifying the origin of the strong phase in the heavy meson decays~\cite{Arnesen08:anni-scet,Chay08:complexanni}.

\item It is interesting to note that the direct and mixing-induced {\it CP} violations are naturally zero in both of the pure
penguin $B_d^0 \to \kappa^0 \bar \kappa^0$ and $B_s^0 \to \kappa^0 \bar \kappa^0$ decays due to lack of the interferences from the
tree contributions in the standard model. Of course, these two channels, especially the latter one with a large branching
ratio as $1.55^{+1.28}_{-0.86} \times 10^{-5}$, could provide a promising platform to test the possible new physics beyond the standard model.
\end{itemize}

\bigskip

In summary, we have studied the {\it CP}-averaged branching ratios and the
{\it CP}-violating asymmetries of the $B \to \kappa \bar \kappa$ decays in
the PQCD approach based on the $k_T$ factorization theorem. The underlying
structure of the light scalars are not determined unambiguously yet. Therefore,
the light scalar $\kappa$ was assumed as a lowest-lying $q\bar q$ meson
in the present work. It is
expected that the productions of the light scalars in the heavy $B$ meson
decays could provide many useful information at another different aspect.
The predictions in the PQCD approach showed that: (1) The large decay rates
above $10^{-6}$ could be found in the $B_d^0 \to \kappa^+ \kappa^-$,
$B_s^0 \to \kappa^+ \kappa^-$, and $B_s^0 \to \kappa^0 \bar \kappa^0$ channels,
which are expected to be measured at the LHCb and/or Belle-II experiments in the
near future; (2) The large direct and mixing-induced {\it CP} violations could be
found in the $B^+ \to \kappa^+ \bar \kappa^0$, $B_d^0 \to \kappa^+ \kappa^-$,
and $B_s^0 \to \kappa^+ \kappa^-$ modes, however, the small branching ratio
$Br(B^+ \to \kappa^+ \bar \kappa^0)$ might limit its future measurements;
(3) The zero direct and mixing-induced {\it CP}-violating asymmetries in the
standard model of the pure penguin $B_d^0 \to \kappa^0 \bar \kappa^0$
and $B_s^0 \to \kappa^0 \bar \kappa^0$ decays would provide a promising platform
to search for the possible new physics beyond the standard model once the nonzero
{\it CP} violations could be detected evidently in these two modes; (4) The QCD
dynamics of the light scalar $\kappa$
is different from that of the $S$-wave pseudoscalar $K$ and vector {\it $K^*(892)$} mesons,
which turned the previously destructive effects into the presently constructive
ones in the nonfactorizable emission and annihilation diagrams, consequently
led to the large branching ratios.

\begin{acknowledgments}
X.L. thanks Prof. Hai-Yang~Cheng for valuable discussions.
This work is supported in part by the National Natural Science
Foundation of China under Grant Nos.~11765012 and 11875033,
by the Qing Lan Project of Jiangsu Province (No.~9212218405), and by the Research
Fund of Jiangsu Normal University (No.~HB2016004).
\end{acknowledgments}


\end{CJK*}
\end{document}